# Personal Data Protection in AI-Native 6G Systems

Keivan Navaie, *Senior Member, IEEE*

*Abstract*— **As 6G emerges as an AI-native technology, seamlessly integrating artificial intelligence (AI) and Generative AI into cellular communication systems, it promises transformative advancements in connectivity, network optimization, and personalized services. However, this integration brings significant data protection challenges as AI models increasingly rely on vast amounts of user data for training, performance optimization, and service personalization. In this context, compliance with stringent data protection regulations, such as the GDPR, is not only necessary but pivotal to the design and operation of 6G networks. These regulations shape critical aspects of system architecture, including transparency, accountability, fairness, bias mitigation, data security, and minimization. This paper identifies the key data protection risk factors associated with the use of AI in 6G and examines how personal data is processed across the 6G lifecycle. We provide a comprehensive analysis of the inherent privacy risks and propose mitigation strategies to address these challenges. Our findings emphasize the urgent need for robust privacy measures that protect user data and uphold individual rights. We argue that integrating privacy-by-design and privacy-by-default principles into the development of 6G standards is imperative for ensuring compliance and fostering trust. Ultimately, a holistic, end-to-end approach to data protection must be adopted, one that identifies critical pain points, assesses the personal information at each stage, and recommends effective data protection techniques for a secure and privacy-conscious 6G ecosystem.**

## I. INTRODUCTION

Building on the foundations established by previous generations, 6G networks promise ultra-reliable, low-latency communication, ubiquitous connectivity, and seamless integration with emerging technologies such as artificial intelligence (AI), the Internet of Things (IoT), and edge computing. The broad scope of 6G, characterized by heterogeneous radio technologies, RAN softwarization, and AI-driven network management, introduces new data protection challenges that affect not only end-users but all stakeholders across the network ecosystem [1].

In 6G systems, personal data flows within the network, serving as a critical enabler for machine learning and AI applications that enhance intelligent network operation. As an AI-native technology, 6G leverages vast amounts of user data to optimize network performance, predict user behavior, and provide personalized services. AI is thus deeply embedded in the 6G landscape, driving intelligent resource allocation, network management, user profiling, and targeted advertising, thereby reshaping the cellular environment.

However, the processing of personal data in these systems is subject to strict regulatory oversight in most countries. Recognizing data protection and privacy as integral elements of the network architecture is crucial, necessitating the development of comprehensive frameworks that address the complex data protection requirements at every layer of the network stack.

The advent of Generative AI (GenAI) has exacerbated privacy challenges in two key areas. First, personal data collected by 6G networks may be used in the training and fine-tuning of GenAI models, introducing novel data protection risks. Second, 6G networks may rely on GenAI models developed by third-party providers, which may involve the transfer of personal data to external entities. These third-party models pose additional risks, as they are often trained on large datasets scraped from the internet, which may include personal data, thus complicating compliance with data protection laws.

This paper contends that addressing privacy challenges in 6G systems and their extended ecosystem is an urgent priority. As AI models permeate various stages of the cellular network life cycle—from design and optimization to operation and maintenance—ensuring compliance with data protection laws and regulations becomes critical. For brevity, both conventional and generative AI models are referred to collectively as AI throughout this paper.

We explore the multifaceted implications of AI on privacy within 6G systems, analyzing how AI technologies influence each stage of the cellular network life cycle, from initial design and deployment to ongoing operation and evolution. By identifying the privacy risks inherent in AI-driven cellular networks, we emphasize the need for robust privacy-preserving measures to protect user data and uphold individual rights. The pursuit of technological innovation must be balanced with the safeguarding of privacy and fundamental human rights. This paper seeks to illuminate the path forward for building a 6G ecosystem that is both technologically advanced and privacy conscious.

The structure of this paper is as follows: Section II defines personal data within the context of 6G networks and examines how it is processed at various stages of the 6G lifecycle. Section III reviews the relevant prior work. Section IV identifies the specific data protection risk factors in 6G networks. Section V offers a comprehensive discussion of data protection risks, potential mitigation strategies, and methods for evaluating their effectiveness. Finally, Section VI concludes by emphasizing the importance of integrating privacy-by-design and privacy-by-default principles in the development of 6G standards. Although this paper focuses primarily on the General Data Protection Regulation (GDPR),

K. Navaie is a Professor of Intelligent Networks at Lancaster University, Lancaster, LA1 4WA, UK. He is also an Independent Member of the Scientific Advisory Committee at the Alan Turing Institute, UK. (Email: k.navaie@lancaster.ac.uk)

the discussion is also applicable to other privacy frameworks, such as the California Consumer Privacy Act (CCPA) and similar regulations.

## II. PERSONAL DATA IN 6G

### A. Definition of Personal Date

The General Data Protection Regulation (GDPR) defines personal data as any information relating to an identified or identifiable natural person. This includes a wide range of identifiers such as names, identification numbers, location data, channel status information, traffic patterns, and online identifiers. In the context of wireless communication systems, personal data under the GDPR encompasses all data pertaining to an identified or identifiable individual, particularly within the various stages of the wireless system life cycle, including design, operation, resource allocation, customer service, inter-operator relations, roaming, and value-added services.

As 6G networks extend beyond traditional boundaries and engage a wider range of stakeholders, adherence to GDPR principles is crucial to maintain privacy and accountability. Data protection remains a central focus in the European Union's research and development efforts for 6G networks, underscoring its importance as a foundational societal concern within the EU's 6G vision. For example, the integration of high-density antennas in known locations, paired with devices linked to specific user identifiers, may introduce new privacy risks, including data leaks and the potential for re-identification of users.

Current regulations, such as the GDPR, not only govern the management of personal data to safeguard user privacy but also shape various aspects of system design. These regulations influence the need for transparency, accountability, data accuracy, fairness, bias mitigation, data security, data minimization, and storage limitation. Therefore, ensuring compliance with these regulations is critical throughout the entire lifecycle of 6G networks—from initial design and deployment to ongoing operation and maintenance.

A core principle of GDPR is the concept of privacy by design and by default [2], which requires that privacy considerations be embedded into the design and operation of systems from the outset. This principle is particularly relevant to 6G networks, where privacy risks must be mitigated early to enhance end-user trust and system integrity. Transparent data practices, informed consent mechanisms, and strong accountability measures are essential components of effective data protection within 6G networks. These measures necessitate a proactive, user-centric approach to safeguarding privacy. Users must be provided with clear and accessible information about data collection practices, the purposes of data processing, and their rights under the GDPR. At the same time, network operators and service providers must demonstrate accountability by adhering to transparent and responsible data-handling practices.

### B. Personal Data Processing in 6G

Throughout the life cycle of wireless communication systems, personal information encompasses a variety of data points associated with end-users, particularly those using smartphones. These data points play critical roles across different phases of cellular network operations, from design and operation to customer service and beyond. Table 1 illustrates examples of the types of personal data processed at various stages.

In the design phase, personal data such as user preferences, device identifiers (like IMEI or MAC address), location data,

**TABLE I: Examples of Personal Data Processing in the 6G Lifecycle**

| Stage of Lifecycle | Purpose of Use Personal Data | Personal Data |
|---|---|---|
| Design Phase | To inform the design and development of network infrastructure, protocols, and services. | device identifiers (such as IMEI or MAC address), location data, and communication patterns. |
| Operation | To optimize the network, plan capacity and deliver service. | Real-time location data, network usage information (e.g., call logs, browsing history, app usage), device metadata (e.g., device type, operating system version), and radio access network information |
| Resource Allocation | To allocate bandwidth, optimize signal strength, manage quality of service, and exploit edge/cloud computing | User profiles, Device identifiers, channel status, service subscriptions, and traffic patterns |
| Customer Service | To facilitate customer service interactions, including user authentication, account management, billing, and troubleshooting | User profiles, contact details, service usage history, and location data. |
| Inter-Operator Relation | To facilitate seamless connectivity and billing reconciliation across different networks | User profile, billing details, location data, and network usage information |
| Value-Added Services | To analyze user behaviour, preferences, and demographics to tailor services and promotions to individual users. | User behaviour, Network Usage data, User Preferences, and demographics |

and communication patterns inform the development of network infrastructure, protocols, and services. This foundational step ensures that the network is tailored to meet the specific needs of its users. As the system transitions to the operation phase, real-time location data, network usage details (e.g., call logs, browsing history, app usage), and device metadata (such as device type and OS version) become critical for network optimization, capacity planning, and service delivery. These data points allow networks to adjust dynamically and efficiently to user demand.

Similarly, resource allocation depends on personal information to make informed decisions about bandwidth distribution, signal optimization, and quality of service prioritization based on user profiles, service subscriptions, and traffic patterns. This ensures that network resources are allocated efficiently to meet the varying needs of users in real time. When users seek customer service, personal information is accessed to aid in user authentication, account management, billing, and troubleshooting. Information like user profiles, contact details, and service usage history facilitates swift and effective customer support.

Beyond individual user interactions, inter-operator relations involve the exchange of personal information, such as subscriber data, billing details, and usage information, to enable seamless connectivity across different networks. This data exchange is critical for services like roaming, where personal information is required for subscriber authentication, location tracking, and billing settlement between home and visited networks. Such coordination ensures uninterrupted service for users as they move between different geographical areas. Additionally, value-added services rely heavily on personal data, analysing user behaviour, preferences, and demographics to deliver personalized content, targeted advertising, and location-based services. These services aim to enhance user experience and engagement by tailoring offerings to individual needs.

Given the centrality of personal information across these stages, safeguarding it is essential for compliance with data protection regulations, protecting user privacy, and building trust with subscribers. To address the risks inherent in the collection, processing, and storage of personal data, robust data protection measures—including anonymization, encryption, access controls, and data minimization—must be implemented. These technical safeguards, paired with transparent privacy policies, user consent mechanisms, and proactive data governance frameworks, foster transparency and accountability. Ultimately, these strategies empower users to manage their personal information effectively, ensuring a secure and trustworthy wireless communication ecosystem.

### III. DATA PROTECTION IN 6G: PREVIOUS WORKS

The advent of 6G technology significantly increases the risk of individual identification and the extraction of vast amounts of personal data, ranging from health information and behavioral patterns to personal beliefs. By leveraging data from sensors, smartphones, and various network-connected devices, 6G systems can analyze and infer user behavior with an unprecedented level of detail.

With 6G enabling seamless communication between these devices, they are no longer operating in isolation. This interconnectedness allows third parties to capture a broad range of signals and extract user information from multiple sources. Furthermore, the inherent network heterogeneity in 6G systems amplifies data protection concerns, introducing risks that surpass those faced in previous generations. These risks are particularly pronounced when adversaries exploit vulnerabilities within system models or transfer data through less secure protocols.

Existing research often focuses on preserving personal information within specific technological components, such as AI-driven network orchestration or the air interface. However, this fragmented approach overlooks the necessity for a system-wide perspective that addresses cross-cutting data protection issues. A narrow focus on individual components neglects critical aspects like data governance and the sheer abundance and specificity of personal data involved in 6G networks—both of which are essential for developing comprehensive data protection strategies.

Some studies have proposed the use of privacy-enhancing AI techniques to bolster data privacy in 6G networks [3]. While these techniques show promise, they require careful integration into the network architecture's specific functions. A significant challenge lies in balancing privacy with explainability, as many privacy-enhancing AI methods lack transparency, introducing new questions regarding model ownership and data ownership. Deploying decentralized AI models on edge platforms, as suggested by recent studies [4], can limit access to data and models, thereby enhancing privacy. However, such decentralization may not be feasible for larger, more complex models due to resource limitations on edge servers. Moreover, AI models deployed through centralized approaches may inadvertently expose insights about their training data, presenting additional privacy risks.

Other methods, such as homomorphic encryption [5], have also been explored as potential solutions. While homomorphic encryption is effective for AI models with simpler architectures, its applicability to more complex models is limited, as it becomes computationally prohibitive with deeper layers. Similarly, blockchain-based solutions [6] and zero-knowledge proofs [7] have been proposed to manage sensitive data in 6G networks. However, these methods often introduce significant computational and signalling overhead, potentially affecting key performance indicators (KPIs) such as latency, which is critical for 6G.

Additionally, encryption methods—both conventional [8] and quantum-resistant [9]—while offering enhanced security, may increase network traffic and computational costs, further impacting network performance. Previous research has predominantly focused on specific types of information, often neglecting broader aspects of data privacy, and while various privacy-enhancing technologies, such as anonymization, have been suggested for use in edge/cloud computing environments, they only address some of the challenges.

Given the limitations of these fragmented approaches, it is clear that there is an urgent need for a holistic, end-to-end strategy for data protection in 6G networks. This strategy should identify the critical pain points, evaluate the personal information involved at each stage of the network life cycle, and recommend appropriate data protection techniques. By doing so, it will enable the development of a comprehensive risk assessment framework to ensure robust and effective data protection throughout the entire 6G ecosystem.

### IV. DATA PROTECTION RISK FACTORS IN 6G

#### A. Data Abundance and Specificity

The proliferation of interconnected devices and sensors within the IoT ecosystem, facilitated by 6G networks, enables the ubiquitous capture of vast volumes of detailed personal information. This abundance of data introduces significant privacy risks, as sensitive personal data is collected, analyzed, and transmitted across the network without necessarily having adequate safeguards in place. The sheer volume and specificity of the data generated allow AI models to make highly accurate inferences about individuals, intensifying concerns over the potential misuse of this information.

6G networks enable the collection of diverse data types—ranging from location and health information to behavioral patterns and personal preferences—often in real time. The potential for detailed analysis and profiling, if left unchecked, poses serious privacy implications. Without robust protections, individuals' private information can be exposed, manipulated, or exploited, making comprehensive data protection mechanisms essential to mitigate these risks.

#### B. Edge/Cloud AI

The shift towards AI-native architecture in 6G networks [4], where AI is deeply integrated into various network functions, presents new challenges in data protection. AI-driven systems in 6G frequently rely on vast amounts of personal data for training and decision-making, raising concerns about fairness, transparency, data minimization, and compliance with storage limitations. As data is processed across multiple entities—including third-party cloud and edge servers—the risk of privacy breaches increases, especially when dealing with sensitive information.

AI models operating on edge or cloud platforms are particularly vulnerable to privacy risks. The deployment of these models at the network edge exposes them to privacy attacks, while their integration into the air interface adds another layer of complexity to privacy considerations. These AI systems can analyze and inferring highly sensitive personal information, potentially leading to issues such as unauthorized profiling, discrimination, and breaches of privacy regulations.

Furthermore, the outputs and predictions generated by AI models may inadvertently provide insights into the underlying training data, posing risks of re-identification and data exposure. The opaque nature of AI decision-making, often referred to as the "black-box" problem, complicates efforts to ensure transparency and accountability, making it difficult to assess compliance with data protection principles. AI-driven functionalities such as predictive analytics and automated decision-making also raise concerns about individual autonomy and informed consent, as users may not fully understand or control how their data is used.

### C. Data Governance

The heterogeneity of 6G networks demands seamless data flows across various entities, systems, and potentially international borders [1], which complicates data governance. Issues of data ownership, control, and provenance become more critical as personal data is exchanged between different jurisdictions and network operators. The decentralized nature of 6G networks, along with the involvement of numerous stakeholders, heightens the importance of establishing clear frameworks for data governance to ensure accountability and compliance with data protection laws.

Additionally, divergent data protection mechanisms across regions and industries may impede the free movement of data and introduce compliance challenges. Ensuring data veracity and maintaining consistent data protection standards across these diverse entities is crucial to safeguard personal information. Effective governance frameworks must not only address who owns and controls the data but also establish protocols for data provenance to ensure that the origin, accuracy, and integrity of the data can be verified at all times.

## V. DATA PROTECTION RISKS AND MITIGATION STRATEGIES IN AI-DRIVEN 6G SYSTEMS

Employing AI in the lifecycle of cellular networks introduces several data protection risks, which are further amplified in 6G systems due to enhanced network operation and service provisioning capabilities. To address these risks, a combination of technical, procedural, and organizational measures must be implemented. This section identifies the key risks and presents mitigation strategies, alongside quantitative metrics to measure their effectiveness.

### A. Privacy Breaches

AI models in 6G systems may process sensitive user data, such as location information, without explicit user consent, potentially leading to privacy breaches. Data may also flow to third-party edge or cloud providers, increasing the risk of unauthorized access.

To mitigate these risks, strict access controls and encryption mechanisms should be implemented to ensure data security. End-to-end encryption should be used for data in transit, while access controls should limit unauthorized personnel from accessing sensitive data. Additionally, explicit user consent must be obtained before data collection or processing.

The efficiency of these measures can be monitored using metrics such as the Access Violation Rate, which tracks unauthorized access attempts, and the Encryption Success Rate, measuring the proportion of correctly encrypted data. The Consent Capture Rate can also be used to assess the percentage of users who provide explicit consent, ensuring compliance with privacy regulations like GDPR.

### B. Data Misuse

AI models may misuse user data for unauthorized purposes, such as profiling or targeted advertising, without appropriate user consent. For instance, user data might be used to generate personalized content recommendations without clear disclosure.

To address this risk, privacy policies should be transparent and explain how user data is collected, processed, and used. Affirmative consent should be obtained from users before using their data for secondary purposes.

The effectiveness of these strategies can be measured through the Transparency Comprehension Score, which assesses user understanding of privacy policies via surveys, and the Opt-In Rate for Secondary Uses, which tracks how many users explicitly consent to having their data used for purposes beyond the original intent.

### C. Lack of Transparency

AI systems may function as "black boxes," making it difficult for users or regulators to understand how their data is processed, and decisions are made. This lack of transparency undermines user control over personal data.

Mitigating this risk involves employing Explainable AI (XAI) techniques, which make AI decisions more transparent to stakeholders. User-friendly interfaces should also be provided to enable users to access and control their data.

The Explainability Coverage metric can measure the percentage of AI models that utilize XAI techniques, while the User Control Interaction Rate tracks how often users engage with data control features, such as opting out or modifying data settings.

### D. Bias and Discrimination

AI models may exhibit bias or discrimination, leading to unequal outcomes for specific user groups based on sensitive attributes like race, gender, or socioeconomic status. For example, a network optimization model might prioritize certain user groups over others, leading to disparities in service quality.

To mitigate these issues, regular bias audits should be conducted to identify and address bias in AI models. Fairness-aware models and diverse data sampling techniques should be used to ensure equitable treatment of all user groups.

The Disparate Impact Ratio (DIR) can be used to measure fairness by comparing outcomes across different demographic groups. Additionally, the Bias Correction Rate can track the percentage of identified biases that are successfully addressed, ensuring continuous improvement in fairness.

### E. Security Vulnerabilities

AI systems in 6G networks are susceptible to security breaches, such as adversarial attacks, which can compromise the confidentiality and integrity of user data. For instance, a malicious actor could exploit vulnerabilities in an AI-based intrusion detection system to evade detection and launch cyberattacks.

Mitigation strategies include rigorous security testing and penetration testing to identify and remediate vulnerabilities. Multi-factor authentication (MFA) and strong encryption should also be implemented to prevent unauthorized access.

The Penetration Testing Success Rate measures the effectiveness of identifying vulnerabilities during testing, while the Authentication Failure Rate tracks the number of failed or bypassed authentication attempts. Lower failure rates indicate stronger system security.

TABLE II: DATA PROTECTION RISK, MITIGATION STRATEGIES, AND EVALUATION METRICS

| Risk | Mitigation Strategy | Efficiency Measures | Quantification |
|---|---|---|---|
| **Privacy Breaches** | Strict access controls, encryption, explicit consent management | Access Violation Rate, Encryption Success Rate, Consent Capture Rate | (Unauthorized access attempts / Total access attempts) X100 |
| **Data Misuse** | Transparent privacy policies, affirmative consent for secondary uses | Transparency Comprehension Score, Opt-In Rate for Secondary Uses | (Users reporting clear understanding / Total surveyed) X100 |
| **Lack of Transparency** | Explainable AI (XAI), user-friendly interfaces for data control | Explainability Coverage, User Control Interaction Rate | (AI models with XAI / Total AI models) X100 |
| **Bias and Discrimination** | Regular bias audits, fairness-aware models, diverse data sampling | Disparate Impact Ratio, Bias Correction Rate | Disparate Impact Ratio = (Positive outcomes group A / group B) |
| **Security Vulnerabilities** | Rigorous security testing, multi-factor authentication, encryption | Penetration Testing Success Rate, Authentication Failure Rate | (Identified vulnerabilities / Total vulnerabilities tested) X100 |
| **Regulatory Compliance** | Comprehensive data governance, regular compliance audits | Compliance Audit Pass Rate, Non-Compliance Incident Rate | (Audits passed / Total audits) X100 |
| **Data Ownership** | Clear data ownership agreements, privacy-enhancing technologies | Agreement Clarity Score, Anonymization Success Rate | (Clear agreements / Total agreements) X100 |

### F. Regulatory Compliance

Failure to comply with data protection regulations such as GDPR or CCPA can lead to legal and financial penalties for cellular network operators. For example, using AI models to process personal data without user consent may result in violations of these regulations.

To mitigate this risk, network providers should implement comprehensive data governance policies and conduct regular compliance audits. These measures help ensure that the use of AI in cellular networks adheres to relevant data protection regulations.

The Compliance Audit Pass Rate measures the percentage of audits passed without regulatory issues, while the Non-Compliance Incident Rate tracks the number of regulatory violations. A lower incident rate reflects stronger adherence to legal standards.

### G. Data Ownership

In AI-driven 6G systems, uncertainties regarding data ownership may arise, particularly when user data is shared with third-party service providers or used for AI training. For example, a communication provider may share data collected from IoT devices with external AI service providers, leading to potential conflicts over data ownership.

To mitigate this, agreements with third-party providers must clearly define data ownership and usage rights. PET such as data anonymization and pseudonymization should be employed to protect user privacy during data sharing.

The *Agreement Clarity Score* measures how well data ownership and usage rights are defined in contracts, while the Anonymization Success Rate tracks the effectiveness of data anonymization techniques in protecting user privacy when sharing data with external parties.

## V. CONCLUSIONS

This paper underscores the significant impact of data protection regulations on 6G systems, particularly as AI models increasingly utilize user data for performance optimization and personalized services. We have identified key data protection risk factors associated with the use of AI in 6G, including data abundance and specificity, the reliance on edge and cloud computing, and the complexities of data governance regimes. By exploring the flow of personal data throughout the various stages of the 6G lifecycle, we have detailed specific data protection risks and provided potential mitigation strategies.

Our findings emphasize the critical need for robust privacy measures to protect user data and uphold individual rights in accordance with regulations such as the GDPR. This analysis highlights the urgent necessity of integrating privacy-by-design and privacy-by-default principles into the development of 6G standards from the inception. By proactively incorporating these principles at the earliest stages of system design, stakeholders can ensure that data protection risks are thoroughly understood and appropriately addressed.

Integrating these measures from the beginning is essential not only for regulatory compliance but also for building user trust and fostering the responsible advancement of technology. As 6G networks continue to evolve, embedding comprehensive data protection frameworks will safeguard user privacy and support the development of secure, ethical, and innovative communication systems in the rapidly changing technological landscape.

## REFERENCES

[1] C. D. Alwis *et al.*, "Survey on 6G Frontiers: Trends, Applications, Requirements, Technologies and Future Research," in *IEEE Open Journal of the ComSoc.*, vol. 2, pp. 836-886, 2021.

[2] "Official Journal L 119/2016," *EUR-Lex*. Archived from the original on Nov. 22, 2018. [Online]. Available: https://eur-lex.europa.eu.

[3] Z. Yang, *et al.*, "Federated Learning for 6G: Applications, Challenges, and Opportunities," *Engineering*, vol. 8, pp. 33-41, 2022.

[4] K. B. Letaief, *et al.*, "Edge Artificial Intelligence for 6G: Vision, Enabling Technologies, and Applications," *IEEE JSAC*, vol. 40, no. 1, pp. 5-36, Jan. 2022.

[5] H. Li, *et al.*, "Lightweight privacy-preserving predictive maintenance in 6G enabled IIoT," *Journal of Industrial Information Integration*, vol. 39, 2024.

[6] Y. Zuo, *et al.*, "A Survey of Blockchain and Artificial Intelligence for 6G Wireless Communications," *IEEE Commu. Surveys & Tutorials*, vol. 25, no. 4, pp. 2494-2528, 2023,

[7] X. Shen, *et al.*, "Blockchain for Transparent Data Management Toward 6G," *Engineering*, vol. 8, pp. 74-85, 2022.

[8] J. Shi, *et al.*, "Toward Data Security in 6G Networks: A Public-Key Searchable Encryption Approach," *IEEE Network*, vol. 36, no. 4, pp. 166-173, July/August 2022.

[9] "Migration to quantum-resistant algorithms in mobile networks," Ericsson Blog, Feb. 2023. [Online]. [Accessed: Jun. 21, 2024].

[10] European Commission, "What does data protection 'by design' and 'by default' mean?" [Online]. [Accessed: Jun. 10, 2024]